\documentclass[conference]{IEEEtran}
\IEEEoverridecommandlockouts
\usepackage{cite}
\usepackage{amsmath,amssymb,amsfonts}
\usepackage{graphicx}
\usepackage{textcomp}
\usepackage{xcolor}
\usepackage{booktabs}
\usepackage{threeparttable}
\usepackage{subfigure}
\usepackage[ruled,linesnumbered]{algorithm2e}
\usepackage{url}
\def\BibTeX{{\rm B\kern-.05em{\sc i\kern-.025em b}\kern-.08em
    T\kern-.1667em\lower.7ex\hbox{E}\kern-.125emX}}
\begin{document}

\title{A Novel Location Free Link Prediction in
Multiplex Social Networks\\
{\footnotesize \textsuperscript{}}
\thanks{}
}

\author{Song Mei\textsuperscript{*}, Cong Zhen\textsuperscript{\dag}\\
\IEEEauthorblockA{\textit{\textsuperscript{*}Huazhong University of Science and Technology, Wuhan, China} \\
\textit{\textsuperscript{\dag}Wuhan Siwei Tongfei Network Technology Company Limited, Wuhan, China}\\
\textsuperscript{*}To whom correspondence should be addressed: meisong@mail.hust.edu.cn}
}

\maketitle

\begin{abstract}
In recent decades, the emergence of social networks has enabled internet service providers (e.g., Facebook, Twitter and Uber) to achieve great commercial success. Link prediction is recognized as a common practice to build the topology of social networks and keep them evolving. Conventionally, link prediction methods are dependent of location information of users, which suffers from information leakage from time to time. To deal with this problem, companies of smart devices (e.g., Apple Inc.) keeps tightening their privacy policy, impeding internet service providers from acquiring location information. Therefore, it is of great importance to design location free link prediction methods, while the accuracy still preserves. In this study, a novel location free link prediction method is proposed for complex social networks. Experiments on real datasets show that the precision of our location free link prediction method increases by 10 percent.
\end{abstract}


\section{Introduction}
For social networks, Twitter, Facebook and Weibo have recently been leaping forward, which encourages people worldwide to connect with the Internet. The so-termed link prediction is one of the basic techniques of social networks, establish user relationships and keeping the network evolving. From the conventional perspective, link prediction has been dependent of the collection of user information (e.g., location and address book). However, as indicated from recent research conducted by Apple \cite{1}, through the application privacy report, users are enabled to know who their data may be shared with by viewing all third-party domain names contacted by the application. Intelligent tracking prevention is conducive to protecting Safari users from unnecessary tracking. It exploits machine learning on the device to prevent tracking, while allowing the website to function normally. Besides, “intelligent tracking prevention” this year hides the user’s IP address, thereby revealing that they cannot use the user’s IP address as a unique identifier to connect to their activities and develop personal information regarding them. Moreover, users lack privacy awareness, and social media encourages them to disclose their personal location information, thereby enabling them to easily find mutual friends \cite{2}. Accordingly, privacy protection turns out to be a huge challenge.

User privacy protection is of high significance to link prediction. The index-based method is achieved by randomly walking all the paths of two users. Existing studies placed major focus on single hop, whereas the existing prediction methods primarily focus on network topology and fail to identify multi-path network sets. Link loss prediction refers to a critical problem with multiplex networks. The user’s online social behavior can be modeled as a mathematical model to accurately represent the user’s social activities. The mentioned dynamic behaviors are capable of simulating real-world user activities to infer the novel links generated. For instance, nodes representing users and edges representing relationships are capable of representing a simple social network. Furthermore, in a multiplex network, to measure the significance of two users, introduce a standardized weight to calculate the number of friendships with two people, this method increase the accuracy of multiple network links.
\begin{figure}[h]
\centering
\includegraphics[width=1.0\linewidth]{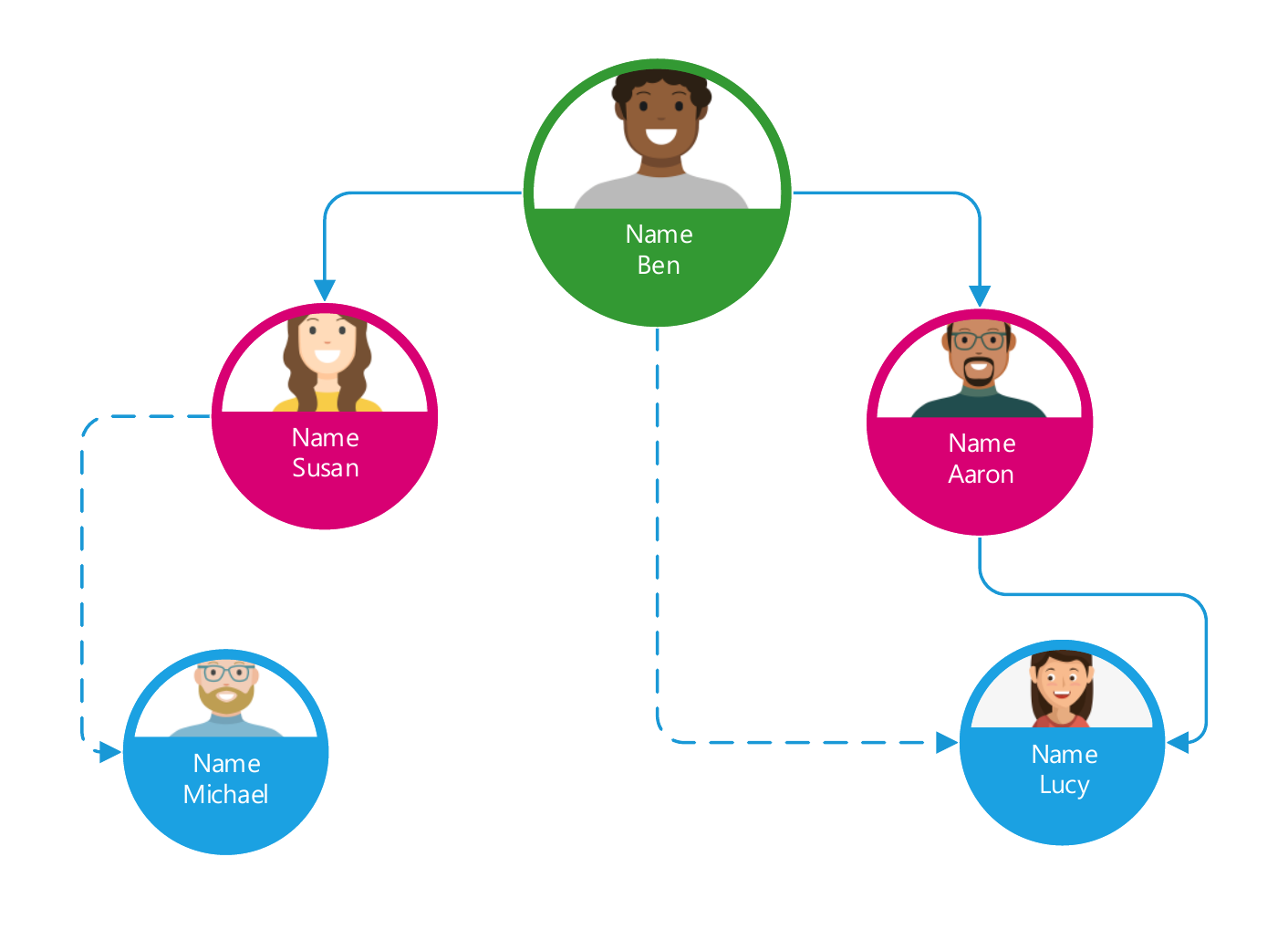}
\caption{Representation of link prediction on social networks}
\label{fig:1}
\end{figure}

The user’s online social activities will generate considerable user information\cite{3}. Through network effect, the behavior of users on social networks is generally impacted by other users \cite{4}. With the rapid development of social networks, more than millions of online users generate a large amount of information on complex data fabric systems. This flood of information brings great challenges to study the dynamic changes in  complex data fabric systems. In fact, in the real complex data structure system, the user interaction between different hot topics leads to the unpredictability of dynamic changes. While current research typically looks at changes in social activity to determine network effects. There by causing link prediction in complicated networks to be more difficult \cite{5}. It is indicated that in complex data fabric systems, network effects cannot well describe the dynamic changing process of social networks. Figure 1 illustrates an example of the target user interacting with another user. In social media, links and social relations between users are considered to be unequal \cite{6}. In a social network, users will have three behaviors of reply, mention and retweet under their friends' blogs or posts. Because the popularity of the topic and the strength of the relationship between friends will affect the user's choice. For example, two users with high intimacy will generate more reply, and ordinary people will mention and retweet more than reply behaviors on posts, which will increase the social influence of mentions and retweet more than reply. This approach makes the reply behavior less influential in the overall behavior ratio, which will eventually lead to the wrong relationship between the prediction task. In order to represent the real social network, we propose a weight method to measure these three behaviors. We assign different weights to different behaviors, instead of determining social influence according to the proportion of the number of network layers. Thus, this study uses a weighted network to reflect the degree of social influence of the relationship. In existing work \cite{7}\cite{8}\cite{9}, weights have also proved to be useful in social relations and user privacy protection.
\begin{figure}[h]
\centering
\includegraphics[width=0.8\linewidth]{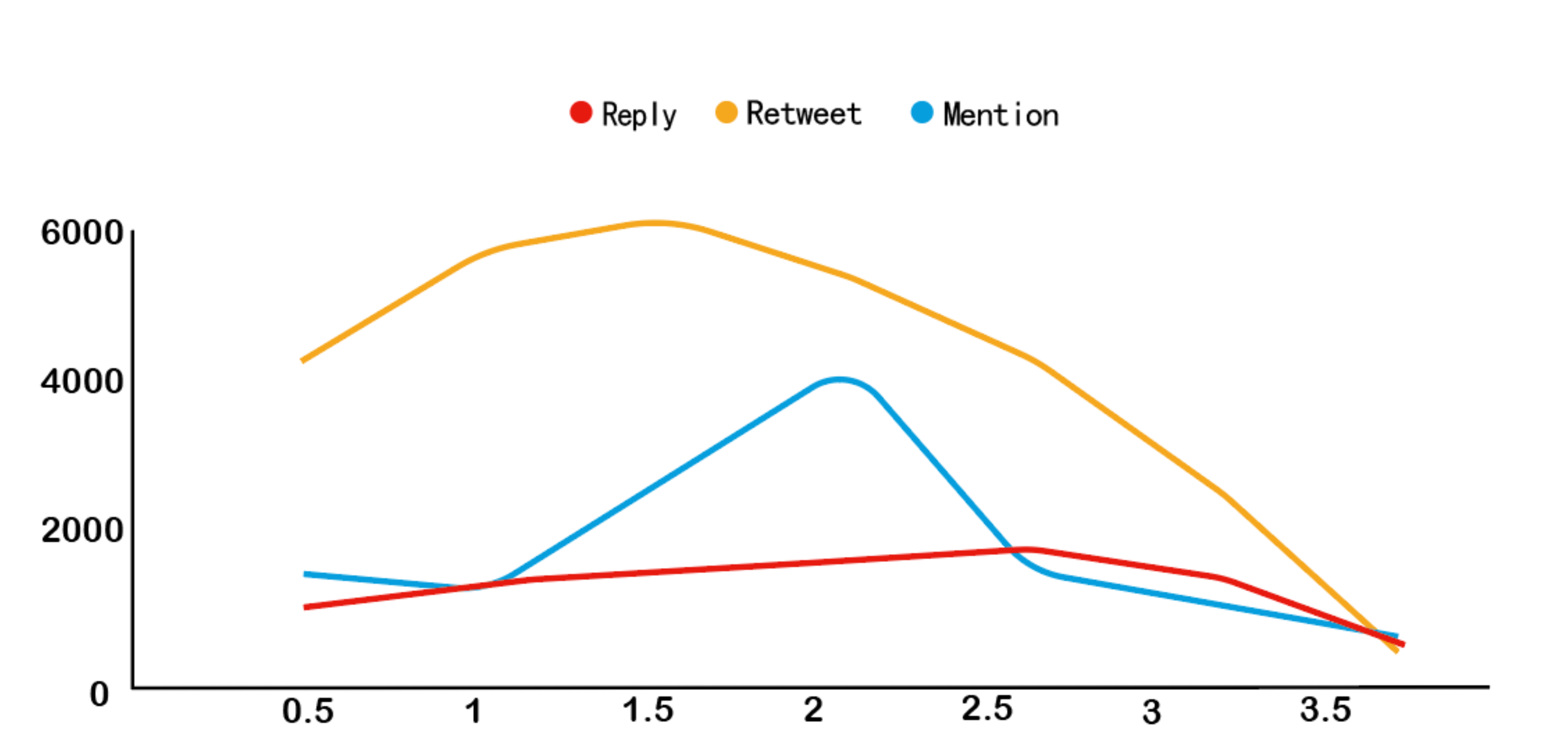}
\caption{Changes in user engagement on topics in social networks over time}
\label{fig:2}
\end{figure}

However, as impacted by the imbalance of user location information, this task turns out to be challenging. First, the correlation between target users and predicted users is heterogeneous. Second, the user’s location cannot be accurately obtained due to the imbalance of location information, which makes the prediction task more difficult. In addition, according to the ML King data. As shown in Figure 2. We selected the network activities of 2500 users in the real social network dataset. By comparing the three behaviors of users in the social network, we took the clustering degree and plotted the lower part of Figure 2. CC represents the correlation clustering of the dataset. The darker the color in Figure 2, the greater the social influence, that is, the higher the engagement to the topic. The abscissa in the lower right figure represents the comparison of the three behaviors, and the ordinate represents the reaction time of the user to make different behaviors per hour. But in order to observe the user’s activity in the social network more intuitively, we use the yellow line, blue line and green line to draw the lower part of Figure 2 into a two-dimensional graph. In the upper part of Figure 2, the yellow, blue and green lines represent the three behaviors of reply (RT), mention (MT) and retweet (RT) generated by users in real social networks. As can be seen from the upper part of Figure 2, users in real social networks are constantly changing with time and topic popularity. Users tend to lower their interest accordingly However, some behaviors (such as reply) do not have a high degree of participation in the network. This part of the data is called the sparse dataset. Because sparse datasets do not occupy a large proportion of the network, they are often filtered out. Conventional methods generally use the structure of the network layer to perform repeated feature extraction at different network layers to collect useful information. Whereas, the link between the less information layer cannot be identified. Due to the complexity and diversity of social networks, traditional forecasting methods do not allow ISPs (Internet Service Provider) to understand the dynamics in social networks. Second, with the introduction of privacy protection policies, data sharing between ISPs is impossible. Under this challenge, ISPs need a new user identity linking algorithm to bring them greater commercial benefits. Accordingly, this study proposes a link prediction method based on non-location information. It starts from the overall situation of the multi-layer network and fully considers the information sparse network layers. The model is capable of truly reflecting the real world.

The main contributions of this study are:

\begin{enumerate}
\setlength{\itemsep}{0pt}
\setlength{\parsep}{0pt}
\setlength{\parskip}{0pt}
\item[(i)] NLFLP(Novel Location Free Link Prediction) model is proposed based on weights by using the concept of significance to define different network layers, and novel link correlations synthesizer is obtained. Thus, the occurrence of false links is reduced.
\item[(ii)] Experiments on multiple real datasets are performed to verify the accuracy and authenticity of the method of different datasets. According to the results, the AUC of this prediction method is 0.1 higher than the baseline.
\item[(iii)] The proposed NLFLP method considers sparse datasets.Experimental results show that in the network layer with relatively sparse data, the prediction of some methods is only 0.16, and our method can also maintain the prediction of 0.75.
\end{enumerate}

\section{Related Work}
In this section, we first give an overview of the existing link prediction work, and then discuss the differences between this research and the above work.

In recently conducted studies, H. Jiang et al.\cite{10} designed a differential privacy model for the problem
of location information leakage. proposed a novel community detection algorithm, that process the fuzzy community structure network. Jiang, H et al. \cite{11} built the signature social network scene on the original
basis, and proposed then the concept of privacy protection. L.Zhao et al.\cite{12} used dynamic routing strategies
to predict the complex space in multiplexed networks. Kou H et al.\cite{13} constructed a hybrid method to
predict the correlation between two adjacent nodes. According to the real social network, the existing
method aims to judge the topological information in accordance with similarity and the likelihood of various
indicators. In other words, it primarily complies with the correlation between information \cite{14}. In the
most natural engineering system, entities are connected with various relationships. In dynamic networks, the
mentioned relationships are critical. The networks can form multiple subsystems and generate various network
information. Tang, R et al.\cite{15} exploited network structure characteristics or network layer information to
address the problem of interlayer prediction. Li bing Wu et al.\cite{16} proposed a bilinear pairing IBEET solution
to solve the matching problem.
P Zhao et al. \cite{17} designed a P3-LOC method to achieve privacy protection for user location information. Existing research mainly adopted the method of network embedding to learn map function. This method unifies the nodes of the network layer into a pre-processing layer for prediction, and it does not affect the position change in the matching node and the common neighbor. However, this method is a static process, failing to truly reflect the real world. L Zhao et al. \cite{18} proposed a method of trajectory prediction based on the generation of confrontation networks on the space problem of multiple networks. Huang S et al. \cite{19} constructed a method based on structural features in the prediction problem of multi-path network. S Wan et al. \cite{20}\cite{21} proposed a layered mobilization method to collect various data packets in a multiplex network. Nevertheless, the mentioned research methods only consider specific examples, thereby failing to achieve the goals predicted in the short-term forecasts, or truly reflect complex social networks.

To perform location free link prediction on a single-layer network, existing scholars have proposed various local and global feature methods. Given the characteristic attributes from the vertex to the edge point, the possibility of missing links is obtained. The method of link prediction complying with the characteristic structure was originally aims at single-layer networks. The mentioned features are termed inter-layer features, and inter-layer features primarily fall to neighborhood-based features and path-based features.
Currently, the common single-layer network prediction methods fall to  different node domain measurement prediction methods and index-based methods. The mentioned methods exploit the presented local domain node information and the characteristics of location attributes for prediction. The common neighbor (CN) method based on two nodes is to calculate the number of adjacent nodes between pairs of nodes, that is, the set A between any pair of nodes, and the length of the path between two nodes is 2. \cite{22}. The Jaccard coefficient (JC) method aims to divide the total nodes in the respective  pair\cite{23}. Adamic-Adar. Adamic-Adar (AA) measurement method is to use the total number of common neighbors and perform reverse weighted search to the identical node. Some relatively simple functional descriptions are presented in the following. The mentioned functions are applied in various link prediction tasks. Moreover, some indicators are briefly explained, which are usually used to measure the similarity between different network layers.

Location free link prediction methods have also achieved effective results over the past few years. H Jiang et al. \cite{24} proposed a differential privacy model to predict the user’s location information. Rodriguez M Z et al. \cite{25} chose a selective parameter clustering method to identify user location data information. Z. Xiao et al. \cite{26} designed a method of regret matching non-cooperative game strategy to improve the accuracy of prediction. proposed a memory network model using similar characteristics of users. Qiu G \cite{27} et al. acquired the user’s location information regarding location services. P Zhao et al. \cite{28} proposed the ILLIA method to solve the privacy protection problem of user location information. However, in the practical world, users often hide their location information by using various methods.  Accordingly, location information is difficult to use for link prediction. H. Jiang et al. \cite{29} designed a Fly-Navi scheme to predict location information. improved the metric to predict links based on the common neighbor algorithm. However, the mentioned research methods only consider specific examples. In short-term predictions, the expected goals are generally not achieved, and they cannot truly reflect complex social networks. As opposed to the mentioned studies, this study focuses on weighted network, differing from the mentioned methods of the focus on the layer’s dynamic and the score aspect of different weights, while avoiding the wrong link.

The research of VP Rekkas et al.\cite{30} summarized the solutions in the case of simple networks. The mentioned problems have been intensively addressed in existing research. Nevertheless, they are too specific to the domain and do not react with real world information. In brief, integrating different levels of information to acquire more complete user information after linking are critical to various ISP. They generally measure the similarity between two nodes or two edges in accordance with different assessment indicators, obtain a relative ranking score of the similarity probability of a node, which is  exploited to infer novel links. \cite{31}\cite{32}\cite{33}. A multi-layer connection method of learning network based on structural features is developed to address the prediction problem of multi-path network. For instance, two adjacent nodes will be affected by nodes of different layers in a multilayer network \cite{34}. To tackle down this problem, Tang F et al. \cite{35} proposed an interlayer network prediction method by embedding vector consistency. Libing Wu et al.\cite{36} proposed a dynamic linear fusion method to predict user ratings. Besides, a novel sorting method is introduced into this study to predict novel links.

\begin{table}
\centering
\newcommand{\tabincell}[2]{\begin{tabular}{@{}#1@{}}#2\end{tabular}}
\caption{Partial similarity measure for link prediction}
  \begin{center}
    \begin{minipage}{\textwidth}
      \begin{tabular}{@{}lll}
        \toprule
        Name & Description & Define \\
        \midrule
        CN &\tabincell{l}{Given two nodes, if the two nodes \\have more  neighbors in common, \\the probability of edge existence is\\ higher. }&		$S(i,j)=\left | \Gamma (i)\cap \Gamma (j) \right |$			\\
        JC &\tabincell{l}{The ratio of the size of the \\intersection of A and B to the size \\ of the union of A and B} &	$	  S(i,j)=\frac{\left | \Gamma (i)\cap \Gamma (j) \right |}{\left | \Gamma (i)\cup \Gamma (j) \right |}$        \\
        LHN & \tabincell{l}{The central suppression index, the\\ closer the node is to the central, \\the less likely it is to have edges.}  & $S_{ij} =\frac{|\Gamma (i)\cap\Gamma (j)|}{k(i)\times k(j) }$       \\
        \bottomrule
      \end{tabular}
    \end{minipage}
  \end{center}
\end{table}

\section{PROPOSED MODEL}
The present section elucidates the NLFLP model. First, the notation and basic
terminology of the model are introduced. Next, various comparison indicators
and the design method of the model are presented. Subsequently, a weighted
link algorithm and ranking method are proposed to reduce the error rate of
prediction.

\subsection{Multiplex Network}
Inconsistent with the conventional single-layer network method, a characteristic of a multi-layer network is that it can flexibly describe various interactions. The connections between nodes in a single-layer network are represented by edges. In addition, nodes and edges constitute a multi-layer network, whereas nodes exist at different layers and represent different forms of interaction. The nodes are linked to each other to develop a network relationship \cite{37}. Various relationships are represented by a range of layers. Besides, the edges of nodes at the identical layer are termed interlayer links \cite{38}. Nodes representing the identical relationship at different layers can be connected by the edges of the layers. It is one of the link prediction problems inter multiplex network to draw upon the multiplex network characteristics extracted from the interlayer and the interlayer correlation between the respective node pair at any layer. Various multilayer networks can be distinguished by the structure of the layer edges. The nodes representing the identical pair of entities \cite{39} at different layers are connected by the edges of the respective layer, and the network falls to a multi-path network.

\subsection{Link prediction in multiplex networks}
Set the graph $G=(V;E;L;C) $ as the multiplexing network composed of nodes.$ V $ denotes the undirected edges between nodes, $ E $ represents the set of edges, $ L$ expresses the labels before different layers, and $ C $ is the connectivity between nodes.$ n $ represents the number of node connections. Set $ U $ as the collection of $ \frac{n(n-1) }{2} $  edges in the respective network layer \cite{40}. Finding the possibility of missing edges at the network layer is the basic problem of link prediction. To find the missing edges, set $ T $ as the set of predicted edges, the edge set  $ E $ is deleted from the original graph, and expresses the edge set of the link prediction algorithm.
$ k $ represents the number of occurrences in the test set. The respective  node $ v $ has a specific label $ L(V) $ at each specific layer. If two nodes are connected, the attribute of the label will increase. Each pair of nodes can connect any number of edges $(u,v)$. There may be different node connections between the respective of edges.

The mapping from network layer G1 to G2 can represent a subgraph isomorphism, if each node v in G1 has a corresponding label in G2, and each pair of nodes (u, v) in the network layer is in the corresponding network. There are corresponding relationships in the layers.

Subgraph Isomorphism:

The mapping function $ f : V1 \rightarrow V2 $ is termed SI.

from $ G1 = (V1, E1, L1, C1) $ to
$ G2 = (V2, E2, L2, C2) $ if

$L_{1}(v) = L_{1}(f(v))$ $\forall v \in V_{t} $

$E_{1}(u,v) \leq E_{2}(f(u),f(v))$ $\forall u,v \in V^{2}_{1} $

Subgraph isomorphism explains the connectivity of network layers. In order to complete the prediction task between network layers, we need to find all the possibilities that nodes have links in the network layers and build a candidate set. The probability that there is a link between the layers of the network is to find the minimum candidate set. For network layer G1, the minimum candidate set contains the minimum set of nodes for this node in network layer G2. If there is a minimum candidate set between network layers G1 and G2, the network layers are said to be connected.

Candidate Sets:

Given a template G1 = (V1, E1, L1, C1) and a world G2 = (V2, E2, L2, C2),
for the respective template node,$v \in V_{1}$ find all world nodes corresponding to
v in at least one link.

find $\bigcup_{f\in F(G1,G2)}f(v) $ for each $v\in V_{1}$

\subsection{weighted in multiplex network}
The weighted multiplexing network comprises weighted networks at different layers. To be specific, the node set exhibits cardinality, and the link set is determined by the cardinality of the node set. The multi-layer network is expressed by an adjacency matrix with elements. First, the weighted links between nodes are judged at the layer. During this process, the link weights between different nodes are defined to consider only positive numbers to simplify the complexity of the weighted multiplexing network. This is not a restriction on the weighted network, since the weight of the link generally does not consider the case of negative numbers in numerous researches on the weighted network.

Set multi-layer network $ G=(L1,L2,...,LM) $, where $ S_{i}=L(V;E_{i}) $ denotes a group of link points in a single-layer network, $ V $ represents a group of nodes at the identical layer,$ E_{i}( i=1,2,....,n) $ expresses a collection of $ i $ types. Different from the link prediction of a single-layer network. Predicting the associated links between multiple network layers is the basic task of multiplex networks. The corresponding node pairs are paired in the prediction layer (Figure 3). Expanding in a multiplex network is the simplest link prediction method. This method is first based on the characteristics of a single-layer network structure. As mentioned above, the use of extended methods will cause interlayer information to be ignored. Due to the neglect of interlayer information or the combination of interlayer information and interlayer structure, the loss of node information in the multi-path network will be caused \cite{41}. Literature \cite{42}\cite{43}\cite{44}\cite{45} proposed different methods to solve the loss of cross-layer information attributed to link prediction tasks. However, they did not consider the transmission of wrong information brought about by wrong links, thereby causing wrong information matching between nodes.

\begin{figure}[h]
\centering
\includegraphics[width=0.4\linewidth]{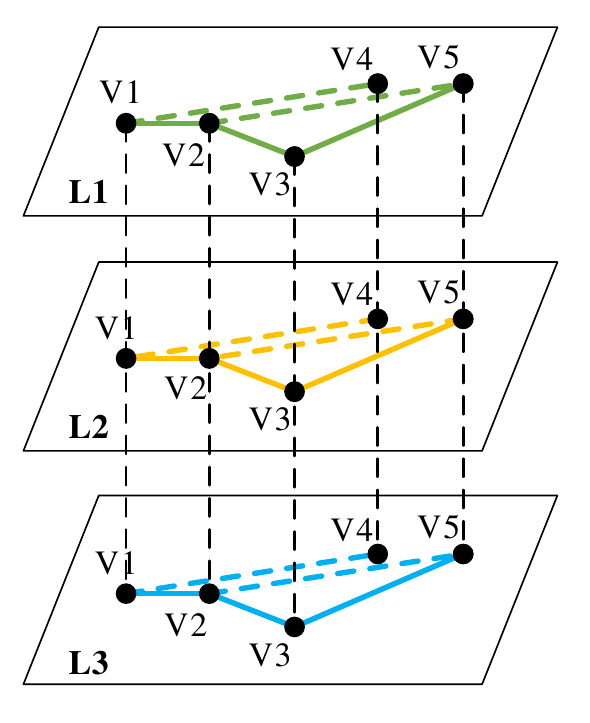}
\caption{Network $ G $  represents a three-layer network containing $ L1, L2, and L3 $. The link prediction problem is to use the inter-layer relationship of the network layer to extract links in a certain layer for prediction.}
\label{fig:3}
\end{figure}

\subsection{Interlayers similarity}
This study proposes using interlayer correlation as the weight of multi-layer network link prediction to enhance the link accuracy in multi-path networks. Though the relevance of multi-layer networks can be measured in various manners \cite{41}\cite{42}\cite{43}, this study only introduces the main relevance of the current research. Set $ G(L1,L2) $ as a two-layer $ L1 $ and $ L2 $ multiplexed network with $ N $ nodes.

\begin{figure}[h]
\centering
 \includegraphics[width=1.0\linewidth]{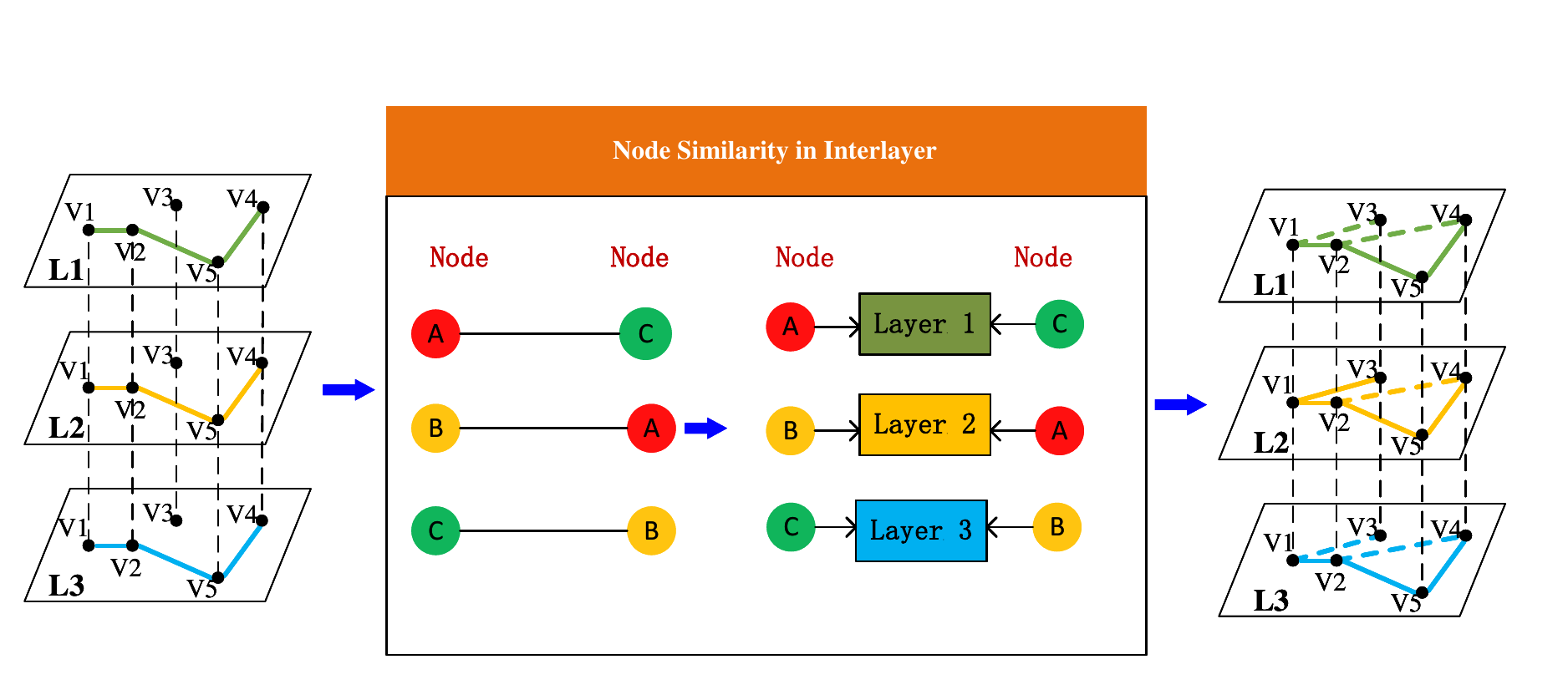}
\caption{A weight-based link prediction framework.}
\label{fig:4}
\end{figure}

\subsubsection{Node centrality}
In network analysis, centrality indicators can determine important nodes at the network layer. Given the obtained node "significance", the obtained ranking can increase the accuracy of our link prediction. "Significance"is generally considered as the closeness of users (nodes) to the network’s participation. Loscalzo J et al. \cite{44} presented a method of centrality measurement to find the central point. The most used method to measure the vitality of a node is to use the centrality of the node, usually it is to count the number of times the node appears on the shortest path. The distance between centrality is defined as:
	\begin{equation}
		D_{CW_{i} } = |CW_{i}^{(L1)} - CW_{i}^{(L2)}|
	\end{equation}	
$ W_{i}^{(L1)} $ is the centrality of node  $ i $ and $ L1 $, then we can define the similarity of node $ i $ between two different layers as:	
	\begin{equation}
		S_{CW_{i}} = 1 - D_{CW_{i}}
	\end{equation}		
	
Through the above definition, we average the layer similarity$ S_{CW_{i}} $ of all nodes, and get the normalized interlayer similarity as:	
	\begin{equation}
			S_{CW} = \frac{\sum_{l=1}^{N}S_{CW_{i}}}{N}
	\end{equation}

\subsubsection{Average similarity of neighbors}
As mentioned above, the average similarity between common neighbors refers to a reciprocal relationship. The easiest way to solve this problem is to standardize the metric and the total number of links at the network layer through network extension, as well as to employ the network density as the metric. The benefit of this standardization is that through the sufficiently high node similarity between layers, the greater the density of this layer, the more information regarding the link prediction contained between it and other layers. Zhao D et al. \cite{46} adopted the asymmetry of average similarity  to propose a novel concept. Given this concept, the mathematics of the similarity between layers is set below:
\begin{equation}
	S_{AASN(L1,L2)} = \frac{\sum_{i}K_{L}(i)}{\sum_{i}K_{L1}(i)}
	\end{equation}	
\begin{equation}
	S_{AASN(L2,L1)} = \frac{\sum_{i}K_{L}(i)}{\sum_{i}K_{L2}(i)}
	\end{equation}
$S_{AASN(L1,L2)}$ denotes the similarity between layers $L1$ and $L2$. In fact, it is predicted that the links in $L2$ will pass this metric and exploit the information in $L1$. Likewise, the similarity between layers of this concept is defined as:
\begin{equation}
	S_{AASN(L1,L2)} = \frac{K_{L}(i)+K_{L}(j)}{K_{L1}(i)+K_{L1}(j)}
	\end{equation}	
\begin{equation}
	S_{AASN(L2,L1)} = \frac{K_{L}(i)+K_{L}(j)}{K_{L2}(i)+K_{L2}(j)}
	\end{equation}	

\subsection{Weighted Prediction Methodology}
As mentioned above, the similarity of the similarity between two nodes is determined by the assigned score of the link prediction algorithm obtained under a specific attribute. The higher the distribution scores achieved by two nodes through calculation, the greater the possibility of correlation between them will be. Thus, the criterion for judging whether there is a potential connection between nodes refers to assigning scores. Given the proposed method, there is a link between two nodes, assuming that node $ x $ and node $ y $ may be related in the future. The link between the two nodes and the edge of their neighbors impacts the size of the weight. Calculate the neighbors with the highest passing scores between the pair of nodes.
Subsequently, get the weight between the selected neighbor sets, and the corresponding link node are weighted and then averaged. Under large differences in the weights between different layers, normalization is performed. Obviously, the more network layers presents more information, the performance of this method is higher when we only consider the non-arbitrary network layers. The specific definition is as follows:

 \begin{equation}
	W_{XY} =\frac{W_{T1}.W_{AX}+W_{T2}.W_{BX}}{2}
\end{equation}
 \begin{equation}
	 W_{T1} =1+\rho \frac{S_{XY} }{S_{AX}}
\end{equation}
\begin{equation}
	W_{T2} =1+\rho \frac{S_{XY} }{S_{BX}}
\end{equation}

The nodes are denoted as $ X $ and $ Y $, and this research focuses on predicting the weights between nodes, the neighbors with the maximal scores are represented by $ A$ and $ B $; $ W_{xy} $ represents the weight of the link edge between nodes $ X $ and $ Y $. $ W_{AX} $ expresses the weight of the edge between node $ A $ and predicted node $ X $, $ S_{AX} $ is the edge score between node $ A $ and predicted node $ X $, and $ W_{T1} $ and $ W_{T2} $ represent the weight of node $ A $ and node  $ B $, respectively. $ \rho=\frac{E1}{E2} $ , $ E $ is the set of edges.

\begin{equation}
	\begin{cases}
		\frac{E1}{E2}>1 & \text{  } x=2 \\
		others& \text{ } x=1
	\end{cases}
\end{equation}

This method exploits information from other layers to perform link prediction on the target layer. The layers other than the target layer are termed prediction layers. The information contained between layers is different, so the degree of effect of different prediction layers on the target layer is different. Some network layers with covering richer information can predict the target layer more accurately, while some layers may exert a poorer prediction effect on the target layer. Because the amount of information contained in the network layer is not the same. According to the degree of information contained in the network layer, the data onto the network layer is divided into different labels. We define these labels as feature information. On the basis of the division, we use the maximum likelihood estimation method to predict the feature information, and obtain a likelihood score, which participates in the final weighted score. The prediction layer finds the link correspondence between the network layers through the score. The feature information between the target layer and the respective prediction layer is extracted through the structural features of the network layer. This information is extracted, and  the most relevant predictor is found to represent the target layer to determine the overall effect. The final link prediction score is determined through the above weighted results. Algorithm 1 uses the overall information of other network layers to distribute the possibility that the link exists in the target layer. For different target layers, the possibilities are determined separately and then used as weights. The possibility of the overall combination is obtained by calculating the existing snapshot image to conduct a separate possibility calculation and accumulating to obtain an estimated probability. On that basis, a prediction layer can be obtained based on the target layer where the link exists.

\begin{algorithm}[h]
\SetAlgoLined
\SetKwFunction{IL}{InitializeDistance}
\SetKwFunction{PL}{PropagateInsertion}
\SetKwFunction{MIN}{Min}
\SetKwFunction{MX}{Max}
\SetKwFunction{TOP}{Top}
\SetKwFunction{Push}{Push}
\SetKwFunction{Pop}{Pop}
\SetKwFunction{Append}{Append}
\SetKwData{Queue}{Queue}
\KwResult{Allocate (target layers score)}
  \For{$each \ target \ x\in L_{p} $  }{
  $Score(y) \le 0$\;
   $ Weight(layer) \longleftarrow  Layer likelihood(x,y)$\;
   $ link \gets x $\;
   $ link \gets L_{t} $\;   }
   \For{$each \ targetedge \ y \in (U - E_{i})\bigcup T  $}{
     \For{$each \ targetlayer \ x\in L_{p} $}{
        $ Score(y) \longleftarrow  Score(y)*PreLayer(x,y)$\; }   }
 \KwRet{}\;
 \caption{Weight prediction algorithm}
 \label{algo:linear}
\end{algorithm}

Algorithm 2 is used to measure the AUC\cite{47} of the proposed method. It employs an iterative method to
extract two edges. To be specific, one edge originates from the training set, and the other edge comes from
the unobserved part not existing in the graph. If the probability of the correctly predicted edge exceeds that
of the unobserved edge, it will increase the probability by comparing the two edges. Algorithm 3 is to detect
the second measure of our proposed method. It sorts the edges in accordance with the distribution scores
obtained, and then checks the total number of edges that are predicted to be correct. The mentioned correct
edges can be employed to verify the accuracy of our ranking scores. LinkPredictorLayer is adopted to collect
information between different layers. If a link in the prediction layer is captured in a snapshot of the network
graph, it will be exploited to assess the existence of the edge of the predictor layer and then the likelihood
estimate will be used to get the target layer to predict the existence of edges. For edges in the predictor
layer, it will get a return value of 1, otherwise it will get a return value of 0. Repeat the operation for all
layers separately. Earlier network snapshots can be inferred based on network features. After all possible
combinations of the respective prediction layer are obtained, predictions of future links can be conducted.

\begin{algorithm}[h]
\SetAlgoLined
\SetKwFunction{IL}{InitializeDistance}
\SetKwFunction{PL}{PropagateInsertion}
\SetKwFunction{MIN}{Min}
\SetKwFunction{MX}{Max}
\SetKwFunction{TOP}{Top}
\SetKwFunction{Push}{Push}
\SetKwFunction{Pop}{Pop}
\SetKwFunction{Append}{Append}
\SetKwData{Queue}{Queue}
{Require: $Test,AUC \ \alpha,\beta,\gamma$}\;
{Ensure: AUC Score\;
$x \longleftarrow randlayeredge(U-E_{i})$\;
$y \longleftarrow randlayeredge(T)$\;}
\If {$edgeScore(x)>edgeScore(y)$\;
$increment \ \beta $}{
    \Else{
        \If{$edgeScore(x) = edgeSocre(y)$\;
        $increment \ \gamma$ }{}}
}
$increment \ \alpha$\;
$decrement \ number \ of  \ tests$\;
\While{$number \ of  \ test \neq 0$}{
    $AUC = \frac{\beta+\gamma}{\alpha}$}
 \caption{AUC Algorithm}
 \label{algo:linear}
\end{algorithm}

\begin{algorithm}[h]
\SetAlgoLined
\SetKwFunction{IL}{InitializeDistance}
\SetKwFunction{PL}{PropagateInsertion}
\SetKwFunction{MIN}{Min}
\SetKwFunction{MX}{Max}
\SetKwFunction{TOP}{Top}
\SetKwFunction{Push}{Push}
\SetKwFunction{Pop}{Pop}
\SetKwFunction{Append}{Append}
\SetKwData{Queue}{Queue}
\KwResult{obtain (prediciton)}
    $ sortedScore \longleftarrow  Sort(Scores,Descent) $\;
    $ highScore \longleftarrow  Sort(1,\alpha) $\;
    \For{$edges \in highScore$ } {
        \If{$testedge(Scores)\in T$}{
        $increase \ k$\;}
    }
    $Precision = \frac{k}{\alpha}$\;
 \KwRet{}\;
 \caption{Precision Algorithm}
 \label{algo:linear}
\end{algorithm}

\begin{table*}[h]\small
\centering
\newcommand{\tabincell}[2]{\begin{tabular}{@{}#1@{}}#2\end{tabular}}
\begin{threeparttable}
\caption{The respective dataset to represent a special event, and different datasets represent different social events. This study collects different behaviors of different users on Twitter. The multiplex network applied in this study adopts a three-layer structure, respectively corresponding to forwarding, mentioning and replying\label{tab:label}}
\begin{tabular}{@{}llllll}
\toprule
Type    & Name              & Event                     & Layers & Users     & Interactions  \\
\midrule
 Religion & PopeElection2013         & \tabincell{l}{People Francis election in 2013}          & 3    & 2,064,866 & 5,969,189    \\
Sport    & NBAFINALS2015            & NBA finals in 201585                     & 3     & 747,937   & 2,150,187    \\
Sport    & UCLFINAL2016             & \tabincell{l}{UEFA Champions League Final in 2016}      & 3     & 677,145   & 1,673,492    \\
Science  & \tabincell{l}{GRAVITIONAL\\ \_WAVES\_2016} & \tabincell{l}{Gravitational waves discovery in 2016}    & 3     & 362,086   & 721,590      \\
Culture  & Cannes2013               & \tabincell{l}{Cannes Film Festival in 2013}             & 3     & 438,537   & 1,180,173    \\
Culture  & Sanremo2016\_final       & \tabincell{l}{Sanremo Italian Music Festival in 2016}   & 3     & 56,562    & 461,838      \\
Society  & BostonBomb2013           & Boston Attack in 2013                    & 3     & 4,377,184 & 9,480,331    \\
Society  & ParisAttack2015          & Paris Attack in 2015                     & 3     & 1,896,221 & 4,163,947    \\
Society  & MLKing2013               & \tabincell{l}{50th Anniv. of M.L.King's 'I have a dream'} & 3    & 327,707   & 398,230   \\
\bottomrule
\end{tabular}
\end{threeparttable}
\end{table*}

\section{Experimental design}
In order to evaluate the performance of this method in multi-layer networks, we conducted related experiments on real network data sets, and compared the performance with several methods of similarity and weight.

\subsection{Dataset}
In this section, the method to use multi-layer network dataset is briefly introduced to assess our real world. In social networks, billions of people generate content with social technology systems each day, thereby generating a lot of information. However, in considerable information, the complex interaction between individuals and social activities has triggered an explosive activity of collective attention. In the mentioned article \cite{45}, the researcher analyzed hundreds of popular Weibo platforms. Special events can make the overall analysis more representative, so the dataset selected is the major events  occurring on the Twitter platform, from the NBA finals to the pope election and the discovery of gravitational waves. Furthermore, a lot of experiments are performed on two real datasets. The first are the Twitter-Foursquare dataset, courtesy of \cite{48} De Domenico M et al. The mentioned data include trajectory data onto Foursquare (a popular diversified social network) and Twitter (a global Weibo social network). In this dataset, we do not consider the direction of the nodes, there are a total of 2458 anchor pairs left. Another dataset, originates from Facebook, a popular social application of more than 10 million registered users. To construct a multi-layer network dataset, we collect 4225 user comments on Facebook from 2013 to 2016. In addition, to imitate the different social behaviors of users in the real world, we consider the weight of the data itself as a criterion for measuring user behavior.

\subsection{Assessment index}
In order to compare the performance of the method proposed in this paper with other comparison methods, we use some evaluation indicators to measure the performance of the algorithm. In this study, we use two assessment indicators, AUC \cite{47} and Precision.

\noindent\textbf{AUC}\\
We use the AUC indicator to measure the ability of the predictive algorithm to identify positive or negative samples. It is therefore indicated that a link randomly selected from the test set has a probability score greater than or equal to that of the link randomly selected from the training set.
\begin{equation}
	AUC=\frac{\beta +0.5\gamma }{\alpha }
\end{equation}
In the above equation, $ \alpha $ represents the total number of missing edges and non-existent edges, $\beta $ denotes the number of times the missing edges have higher scores than non-existent edges, and $\gamma $ expresses the situation where the scores of the two are the same. A model with a higher AUC score indicates a better effect.

\noindent\textbf{Precision}\\
Another indicator to measure the model is to use the accuracy of predicting missing edges to determine the accuracy. The calculation method is as follows:
\begin{equation}
	Precision=\frac{k }{a}
\end{equation}
Clearly, the higher the accuracy, the better the accuracy of the model.

\begin{figure*}
\centering
\subfigure[]
{\label{fig:subfig:a}\includegraphics[width=0.31\textwidth]{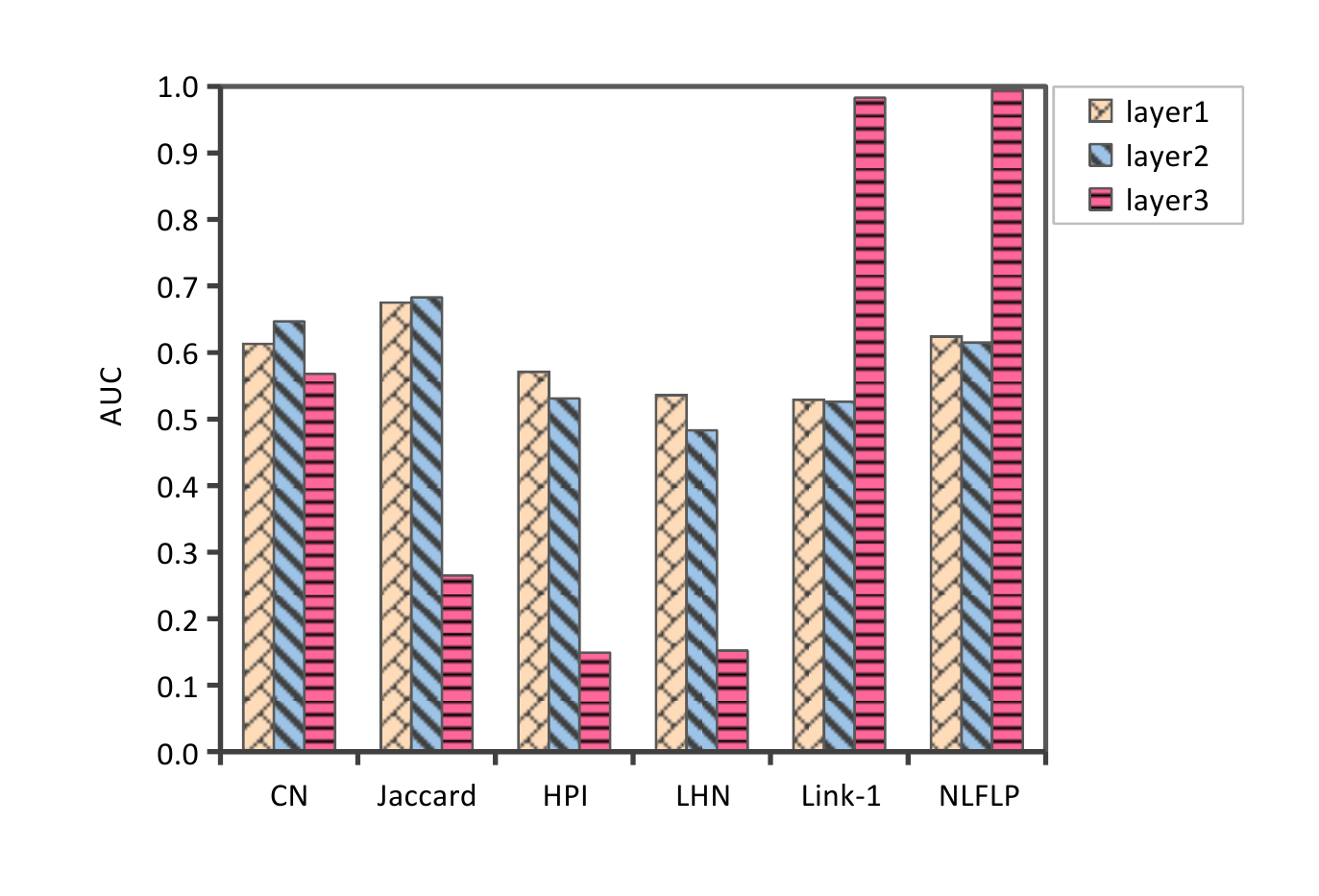}}
\hspace{0.01\linewidth}
\subfigure[]
{\label{fig:subfig:b}
\includegraphics[width=0.31\textwidth]{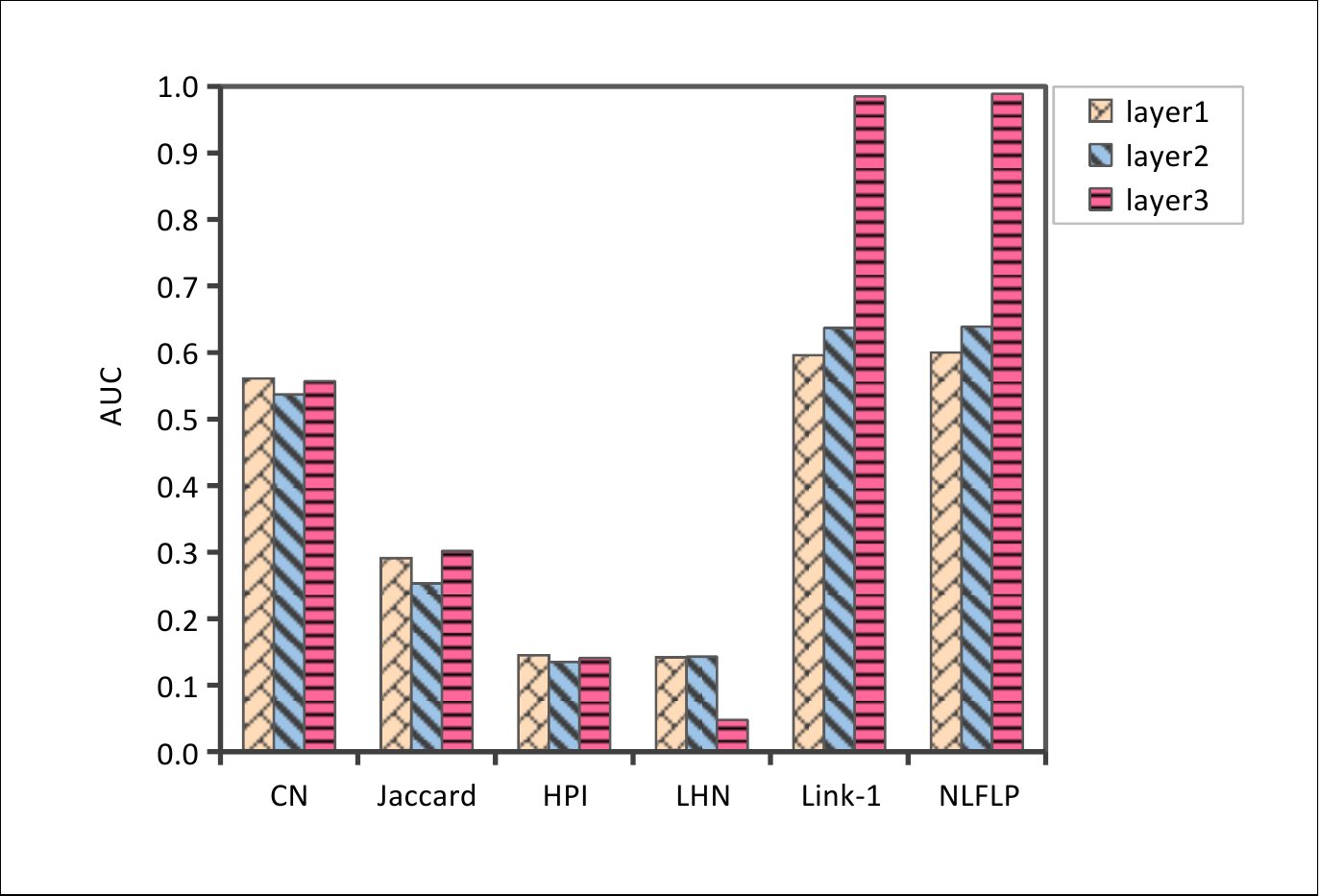}}
\hspace{0.01\linewidth}
\subfigure[]
{\label{fig:subfig:c}\includegraphics[width=0.31\textwidth]{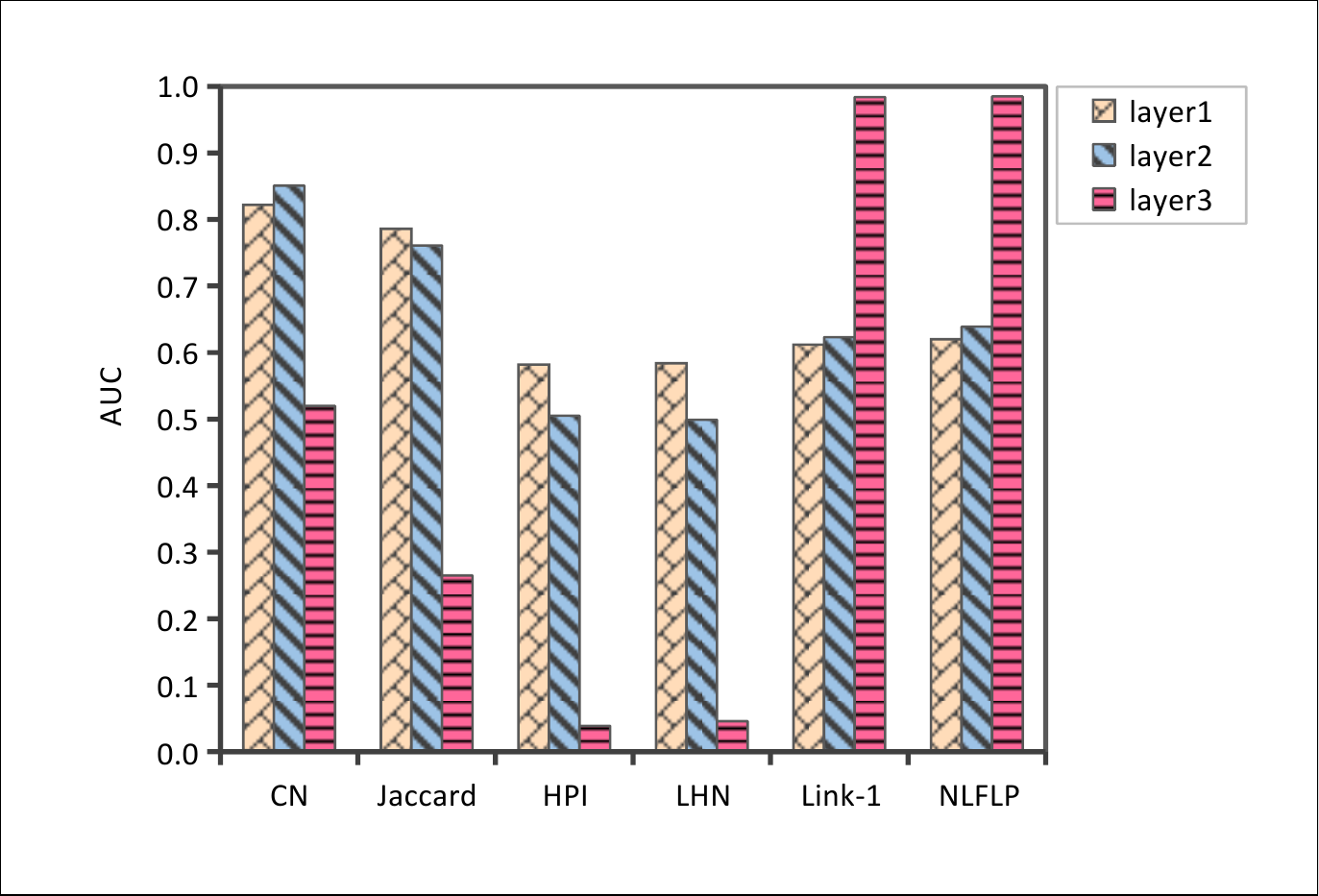}}
\hspace{0.01\linewidth}
\vfill

\subfigure[]
{\label{fig:subfig:d}
\includegraphics[width=0.31\textwidth]{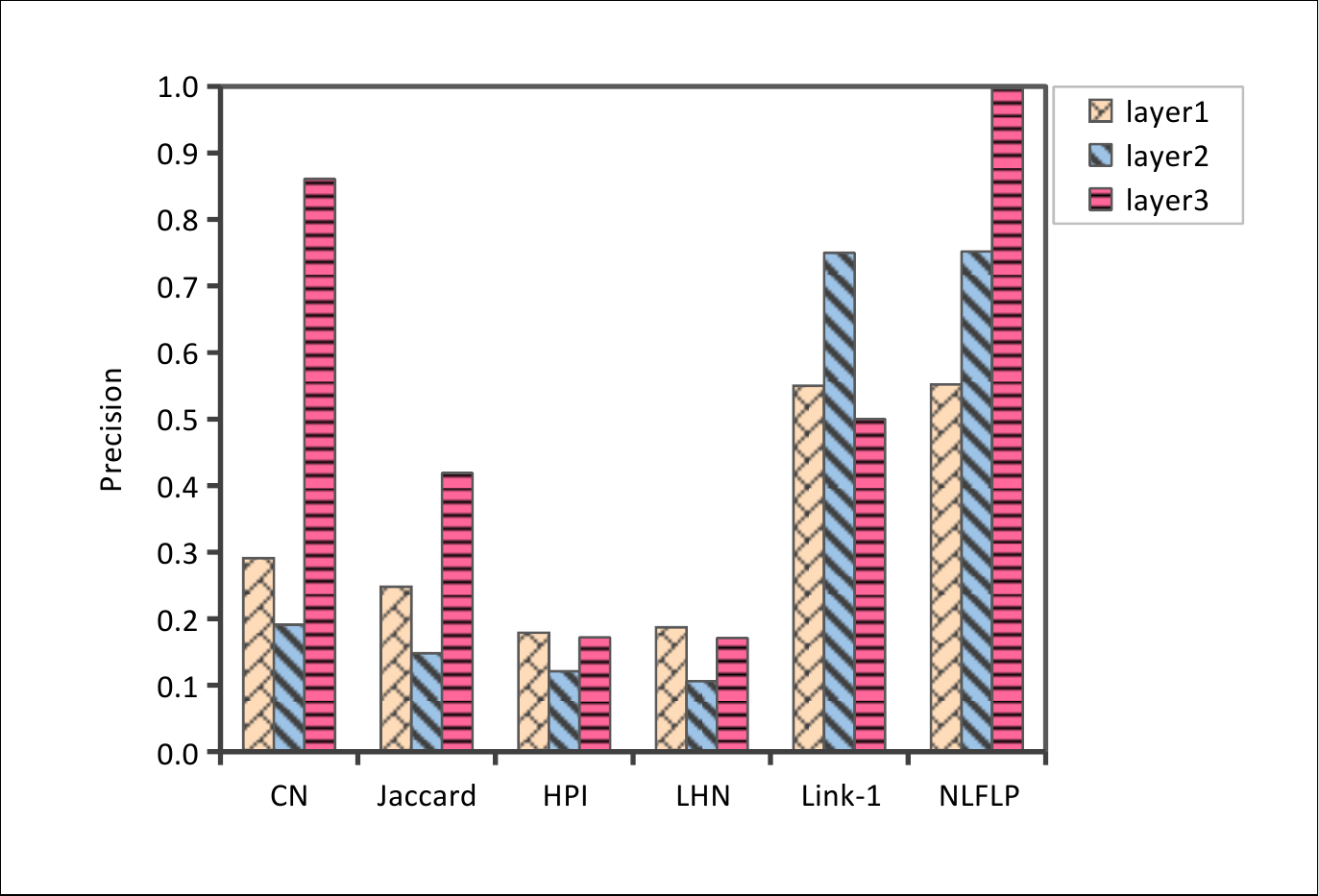}}
\hspace{0.01\linewidth}
\subfigure[]
{\label{fig:subfig:e}\includegraphics[width=0.31\textwidth]{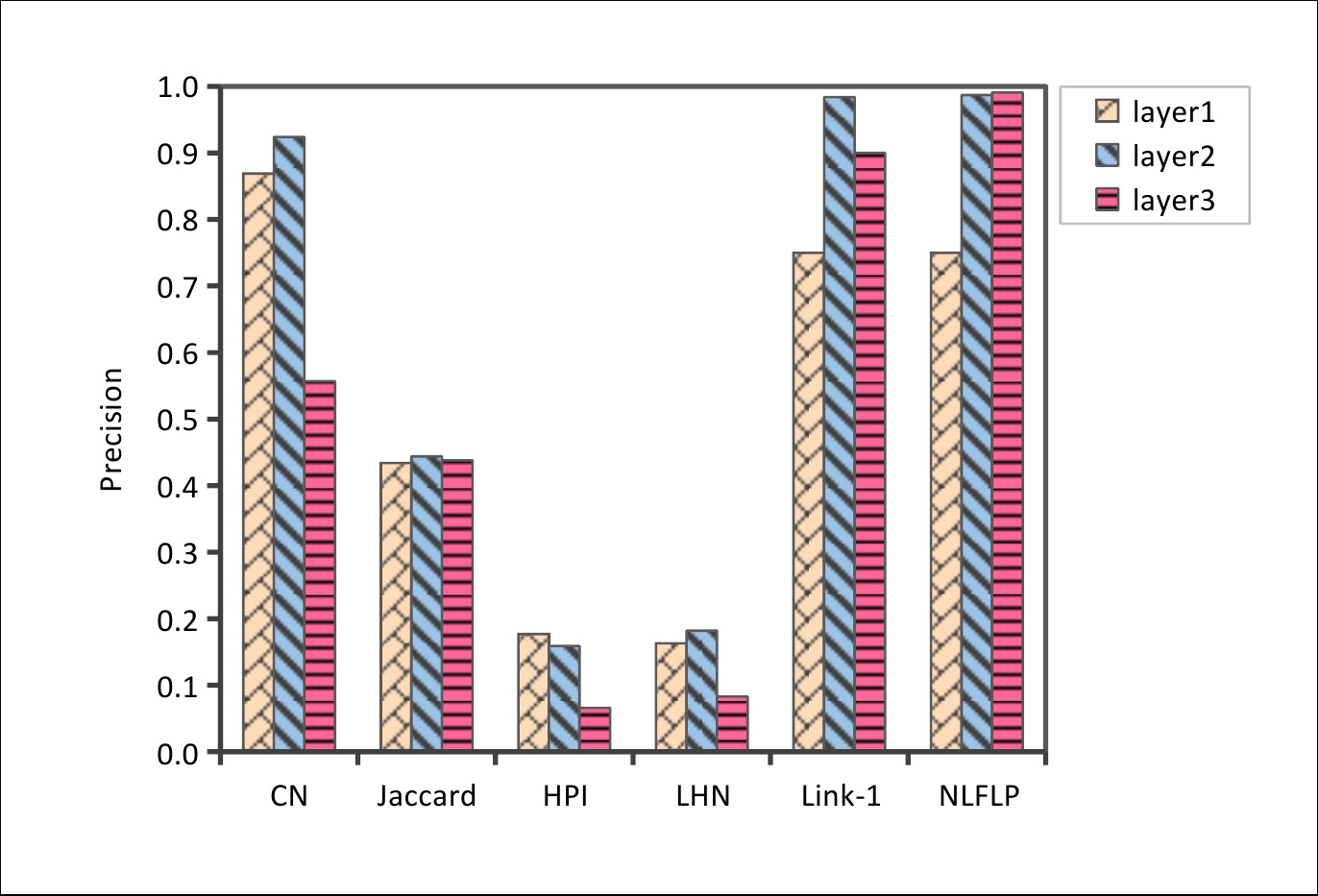}}
\hspace{0.01\linewidth}
\subfigure[]
{\label{fig:subfig:f}
\includegraphics[width=0.31\textwidth]{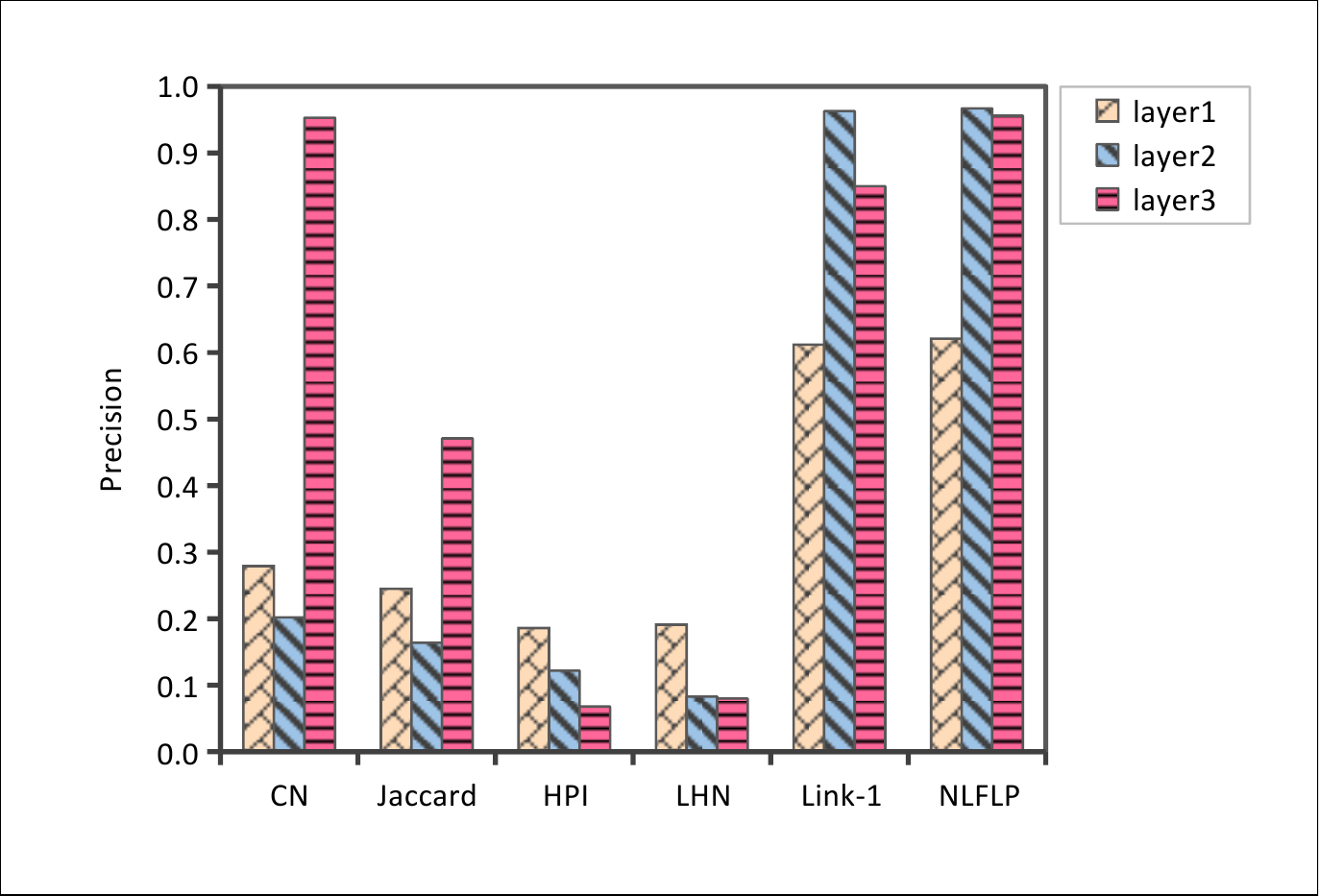}}
\hspace{0.01\linewidth}
\vfill
\label{fig:subfig}
   \caption{(a)(b)(c)Randomly divide 600,800,1000 edges for division, get AUC score.(d)(e)(f)Randomly divide 600,800,1000 edges for division, get precision score.}
\end{figure*}

\subsection{Comparison methods}
In order to compare our method with existing methods, we find the following benchmarks. The present section will describe the mentioned baselines. In order to perform link prediction on a single-layer network, existing scholars have proposed a variety of different locals and global feature methods. The probability of link existence can be obtained through the feature attributes from vertex to edge point. The method of link prediction based on the characteristic structure is originally aimed at single-layer networks. Currently, the common single-layer networks prediction methods are fall to different node domain measurement prediction methods and index-based methods. The mentioned methods use the provided local domain node information and use the characteristics of location attribute to predict. Given the common neighbor between two nodes (CN, common neighbor) method is to count the number of identical neighbors between two points. In other words, the set an of arbitrary connections between the nodes, the path length of the two nodes is 2\cite{37}. The Jaccard coefficient (JC) method is to divide the total number of public neighbors by the total number of neighbors for the respective pair of nodes\cite{38}. Adamic-Adar (AA) measurement method is to use the total number of common neighbors and perform reverse weighting to find the same node. In table 1, We have carried out some relatively simple functional descriptions. The mentioned functions are used in various link prediction tasks. Furthermore, some indicators are briefly explained, which are usually adopted to measure the similarity between different network layers.

\subsection{Experimental results}
To fully to prove that our proposed method is better, our experiment falls to two parts, the first part is to divide the data into the target layer and the prediction layer. The second part is to filter the data. Then, the performance is compared with the conventional five methods and the method of \cite{24} Sharma S et al. Finally, proves that the method is effective against the coefficient dataset.

\subsubsection{results}
The method mainly goes through two steps. The task of the previous step is to obtain the target layer, and the determination of the target layer is determined by the prediction layer. The target layer is a criterion for judging the quality of the model. The next task is to segment the data set through the prediction layer. In this experiment, we randomly select 10 percent in the prediction layer, and the selected part is called the edge set. The remaining 90 percent constitutes the test set. The next task is to segment the data set through the prediction layer. To avoid accidents, the whole process were repeated 20 times.

De Domenico M et al.\cite{48} collected ten different social activity datasets (in the original paper, representing the blog comments made by users). User similarity was evaluated by user tags attributes \cite{49}. Their research results show that they find the node pair with the highest score of each node, and then perform a weighted average between the candidate neighbor set and the corresponding node. All the node data onto different layers are applied, representing the number of users in their papers. The data onto the remaining users are at different network layers. The study compares the similarity scores of different network layers with several others conventional methods to assess the superiority of their proposed methods. Furthermore, other conventional methods of the experiment are added.

Figure 5 presents the standardized privacy score calculated with five methods. As indicated from the results, the proposed method outperforms CN algorithm, Jaccard measure and HDI algorithm in both assessment indexes. At different network layers, the proposed method exhibits better AUC and accuracy than other methods. At the network layer where the data is relatively sparse, and some methods are only 0.16, this method can also maintain an accuracy of 0.75. On the other hand, when the dataset is rich in information, the proposed method also achieves an accuracy of 0.98, while other methods only have 0.8 and 0.6. Many tests were performed on the social event dataset, and 600, 800, 1000 groups of different edges were randomly divided for experiments. Moreover, the accuracy is compared with the method proposed to \cite{49} at the network layer with sparse information.As indicated from the result, the proposed method can also achieve higher performance in sparse datasets \cite{50} \cite{51}\cite{52} \cite{53}\cite{54} \cite{55} \cite{56}\cite{57} \cite{58}.

\section{Conclusion}
We solve the problem of free links to social networks without location. It starts from the overall situation of the multi-layer network, and fully considers the network layer with less information to be filtered. This algorithm establishes the possibility of the existence of edges of layers. The operation is repeated for all layers to determine the individual likelihood scores of different layers. Subsequently, different levels of information are used to give the degree of significance to make the final judgment. It is concluded experimentally that the algorithm outperforms public neighbor, Jaccard's, HDI and other algorithms. It has been verified from two different indicators (AUC, Precision), and the results reveal that the NLFP method achieves better results. The algorithm successfully exploits the information at different layers of the multipath network. Since they encapsulate various heterogeneous correlations between different layers, instead of only a single correlationship, better prediction results are achieved. Moreover, a novel likelihood allocation algorithm is proposed to link prediction. Experimental analysis is conducted on 6 real datasets. Finally, it is proved that the method can calculate the layer with less information and has better accuracy.

\section*{Acknowledgements}
This work was supported in part by the National Natural Science Foundation of China [62002104] [62102136], the 2020 Opening fund for Hubei Key Laboratory of Intelligent Vision Based Monitoring for Hydroelectric Engineering [2020SDSJ06] and the Construction fund for Hubei Key Laboratory of Intelligent Vision Based Monitoring for Hydroelectric Engineering [2019ZYYD007].

\bibliographystyle{unsrt}
\bibliography{conference_101719}

\end{document}